\documentclass[pre, superscriptaddress,showpacs,twocolumn]{revtex4-1}

\usepackage[utf8]{inputenc}

\usepackage{amsmath}
\usepackage{amsfonts}
\usepackage{amssymb}
\usepackage{graphicx}

\begin{document}

\title{Geometric origin of negative Casimir entropies: A scattering-channel analysis}
\author{Gert-Ludwig Ingold}
\email{gert.ingold@physik.uni-augsburg.de}
\affiliation{Institut f{\"u}r Physik, Universit{\"a}t Augsburg,
     Universit{\"a}tsstra{\ss}e 1, D-86135 Augsburg, Germany}
\author{Stefan Umrath}
\affiliation{Institut f{\"u}r Physik, Universit{\"a}t Augsburg,
     Universit{\"a}tsstra{\ss}e 1, D-86135 Augsburg, Germany}
\author{Michael Hartmann}
\affiliation{Institut f{\"u}r Physik, Universit{\"a}t Augsburg,
     Universit{\"a}tsstra{\ss}e 1, D-86135 Augsburg, Germany}
\author{Romain Guérout}
\affiliation{Laboratoire Kastler Brossel, UPMC-Sorbonne Universit{\'e}s,
     ENS-PSL Research University, CNRS, Coll{\`e}ge de France,
     4 place Jussieu, Case 74, 75005 Paris, France}
\author{Astrid Lambrecht}
\affiliation{Laboratoire Kastler Brossel, UPMC-Sorbonne Universit{\'e}s,
     ENS-PSL Research University, CNRS, Coll{\`e}ge de France,
     4 place Jussieu, Case 74, 75005 Paris, France}
\email{astrid.lambrecht@lkb.ens.fr}
\author{Serge Reynaud}
\affiliation{Laboratoire Kastler Brossel, UPMC-Sorbonne Universit{\'e}s,
     ENS-PSL Research University, CNRS, Coll{\`e}ge de France,
     4 place Jussieu, Case 74, 75005 Paris, France}
\author{Kimball A. Milton}
\email{kmilton@ou.edu}
\affiliation{Laboratoire Kastler Brossel, UPMC-Sorbonne Universit{\'e}s,
     ENS-PSL Research University, CNRS, Coll{\`e}ge de France,
     4 place Jussieu, Case 74, 75005 Paris, France}
\affiliation{H. L. Dodge Department of Physics and Astronomy,
     University of Oklahoma, Norman, OK 73019, USA}

\date{\today}

\begin{abstract}

Negative values of the Casimir entropy occur quite frequently at low
temperatures in arrangements of metallic objects. The physical reason lies
either in the dissipative nature of the metals as is the case for the
plane-plane geometry or in the geometric form of the objects involved. Examples
for the latter are the sphere-plane and the sphere-sphere geometry, where
negative Casimir entropies can occur already for perfect metal objects. After
appropriately scaling out the size of the objects, negative Casimir entropies
of geometric origin are particularly pronounced in the limit of large distances
between the objects. We analyze this limit in terms of the different
scattering channels and demonstrate how the negativity of the Casimir entropy
is related to the polarization mixing arising in the scattering process. 
If all involved objects have a finite zero-frequency conductivity, the channels
involving transverse electric modes are suppressed and the Casimir entropy 
within the large-distance limit is found to be positive.

\end{abstract}


\maketitle

\section{Introduction}

The renewed interest in the Casimir effect \cite{Casimir1948} in the last 15
years has fueled numerous complementary studies on different aspects of this
long-range interaction. During the last decade or so various papers have
focused on the entropy related to the Casimir interactions between two solid
bodies \cite{Bezerra2002,Bostroem2004,Bezerra2004,Brevik2004,Brevik2005,
Svetovoy2005,Brevik2006,Bezerra2008,Ellingsen2009,Ingold2009,Pitaevskii2010,
Canaguier2010,Bordag2010,Weber2010,Zandi2010,Rodriguez2011}. An important
reason for this interest lies in the fact that the Casimir entropy can reach
negative values. This effect has first been theoretically discussed for two
plates modelled by Drude-type metals \cite{Bezerra2002}. Its origin lies in the
suppression of the reflectivity of the transverse electric mode at low
frequencies due to a finite zero-frequency conductivity \cite{Bostroem2004}. If
the conductivity diverges in the low-frequency limit, as is the case for
perfect metals or metals within the plasma model, the entropy in the
plane-plane configuration will remain positive for all temperatures
\cite{Bezerra2004,Brevik2005}. The existence of a negative Casimir entropy in
this geometry is therefore related to the dissipation inside the metal.

However dissipation is not the only mechanism giving rise to negative Casimir
entropies. For the Casimir-Polder interaction between an atom and a
perfect-metal plate, negative values of the entropy were found at low
temperatures \cite{Bezerra2008}. Studies of the sphere-plane configuration
\cite{Canaguier2010} and the sphere-sphere configuration \cite{Rodriguez2011}
for perfect reflectors demonstrated that negative Casimir entropies can be of
purely geometric origin. Considering the large-distance limit, we shall see
that in these geometries the Casimir entropy can become negative for perfect
metals, but it remains positive if both objects involved are described by the
Drude model implying a finite zero-frequency conductivity.  

Another indication of a negative Casimir entropy is given by the non-monotonic
behavior of the thermal Casimir force. Such a behavior has been found in the
electromagnetic case in the sphere-plane geometry \cite{Zandi2010} as well  as
for a scalar field satisfying Dirichlet boundary conditions in the sphere-plane
and the cylinder-plane geometries \cite{Weber2010}.

In the present paper we will study in more detail the geometrical origin of
negative Casimir entropies. Before doing so, we emphasize that negative values
of the Casimir entropy are not in conflict with well established thermodynamic
principles. While entropies should be positive, the Casimir entropy actually is
a difference of entropies and can therefore well be negative
\cite{Pitaevskii2010}. However, even the Casimir entropy has to tend towards
zero in the zero-temperature limit. This is indeed the case for the Drude model
for any nonvanishing value of the zero-frequency conductivity
\cite{Brevik2004,Ingold2009}.

Furthermore, close inspection reveals that strictly positive Casimir entropies
over the whole temperature range are rather the exception than the rule.  In
order to explore the geometric origin of the negative Casimir entropy, a recent
study considered the retarded Casimir-Polder interaction between a nanoobject
and a plane or between two nanoobjects \cite{Milton2014}. The properties of the
nanoobjects were described in terms of their electric and magnetic
polarizabilities. By allowing also for anisotropic polarizabilities, a variety
of scenarios could be generated, thus providing insight into the conditions
under which negative Casimir entropies can occur.

Here, we start from a scattering approach for the electromagnetic field and
emphasize the contributions of different scattering channels. First, we
observe that the negativity of the Casimir entropy becomes most pronounced
when the distance between sphere and plane or between the two spheres is
large compared to the radius of the spheres. This observation holds despite
the fact that the entropy scales with the third power of the radius for small
radii so that its overall value will be strongly suppressed. As a consequence,
the large-distance limit is appropriate to identify the physical mechanism
responsible for the negative Casimir entropy. In particular, this limit permits
us to restrict the reflection at the spheres to the dipole modes, $\ell=1$.

The large-distance limit considerably simplifies the expression for the
Casimir free energy within the scattering approach as we shall see in
Sect.~\ref{sec:largeDistanceApproximation}. Firstly, the values of the quantum
number $m$ of the $z$-component of the angular momentum, which is conserved
for geometries of interest here, are constrained to $\vert m\vert=0$ and $1$.
Secondly, it is sufficient to account for one single scattering roundtrip
between the two objects involved. In total, we are left with three distinct
types of channels. Two of these channels leave the polarization type unchanged
and are only distinguished by the value of $\vert m\vert$. The third channel
involves a change of polarization and occurs only for $\vert m\vert=1$.

The most relevant channel for the negative Casimir entropy is the last one. As
we shall see, this scattering channel is characterized by a Casimir free energy
which increases monotonically with temperature. Its contribution to the Casimir
entropy vanishes at zero temperature as well as in the high-temperature limit,
but is negative for any temperatures in between. This result points towards the
importance of polarization mixing in the appearance of a negative Casimir
entropy.

The scenarios of different geometries and zero-frequency conductivity sketched
in the beginning of this section fit nicely into this picture. In the
plane-plane configuration, the polarization is conserved at each reflection.
Therefore, the negative Casimir entropy appearing for Drude-type damping in
that case cannot be of geometric origin. On the other hand, for the
sphere-plane and sphere-sphere configurations, Drude-type metals will suppress
polarization mixing. Therefore, it is to be expected that at least one of the
two scatterers should have a divergent zero-frequency conductivity in order to
allow for a negative Casimir entropy in the large-distance limit.

\section{Negative Casimir entropy in the sphere-plane and the sphere-sphere
geometries}

\begin{figure}
 \begin{center}
  \includegraphics[width=\columnwidth]{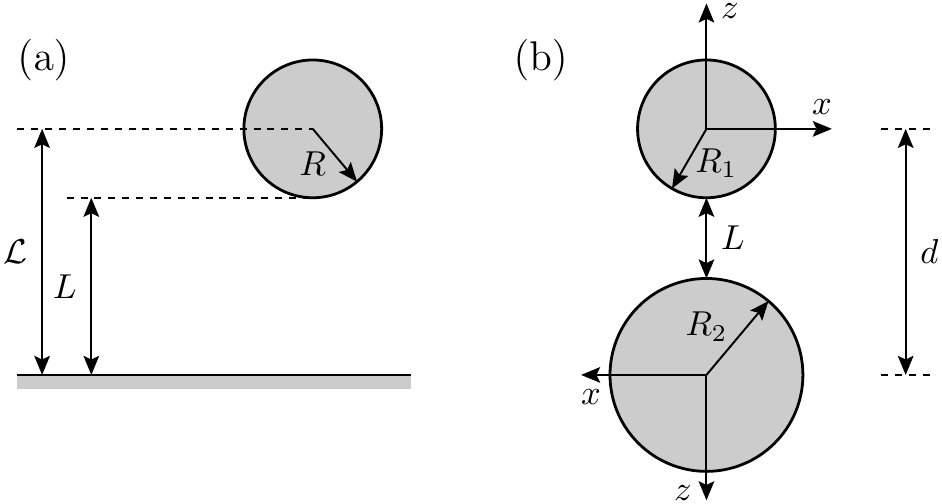}
 \end{center}
 \caption{(a) Plane-sphere and (b) sphere-sphere geometry with the
  geometric parametrization used in this paper. The surface-to-surface
  distance is always referred to as $L$ while $\mathcal{L}$ denotes the
  distance between the plane and the center of the sphere and $d$ denotes
  the distance between the centers of the spheres.}
 \label{fig:geometry}
\end{figure}

In our analysis of the negative Casimir entropy, we will concentrate on the
sphere-plane configuration and the sphere-sphere configuration depicted in
Figs.~\ref{fig:geometry}a and b, respectively, together with the corresponding
geometric parameters. The distance between the surfaces of the two objects
will always be denoted by $L$. For the plane-sphere geometry, the natural
length scale within our analysis will turn out to be $\mathcal{L}=L+R$, which
measures the distance between the surface of the plane and the center of the
sphere of radius $R$. In the sphere-sphere geometry, the natural length scale
$d=L+R_1+R_2$ refers to the distance between the centers of the two spheres
with radii $R_1$ and $R_2$.

Whenever we refer to the two geometries at the same time, we will denote the
natural length scales as $D$. For example, as a dimensionless temperature we
will use
\begin{equation}
\label{eq:dimensionlessTemperature}
\nu =\frac{2\pi Dk_BT}{\hbar c}\,,
\end{equation}
which implies $\nu=2\pi\mathcal{L}k_BT/\hbar c$ in the plane-sphere geometry
and $\nu=2\pi dk_BT/\hbar c$ in the sphere-sphere geometry. Here, $k_B$,
$\hbar$, and $c$ are the Boltzmann constant, the Planck constant and the speed
of light, respectively. In this paper, we make use of a formalism based on
imaginary frequencies $\xi$. The corresponding dimensionless imaginary
frequency is defined by
\begin{equation}
\label{eq:dimensionlessFrequency}
\tilde\xi = \frac{\xi D}{c}\,.
\end{equation}

\begin{figure}
 \begin{center}
  \includegraphics[width=\columnwidth]{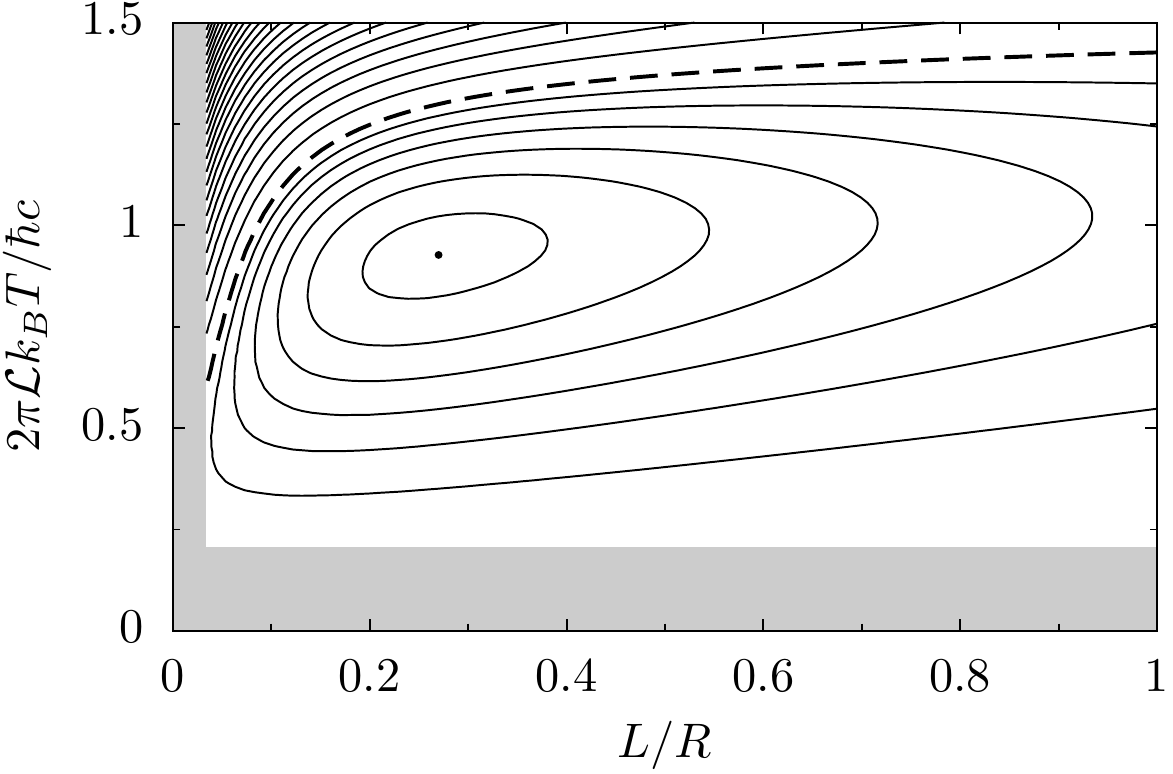}
 \end{center}
 \caption{The Casimir entropy in the sphere-plane geometry is depicted for
  perfect metals as a function of the distance $L$ between sphere and plane and
  the temperature $T$. The entropy vanishes along the dashed line. Below this
  line, the entropy is negative and changes in steps of 0.0005. The minimum of
  the entropy is marked by a dot. Above the dashed line, the entropy is
  positive and changes in steps of 0.001. No data have been calculated in the
  grey region.}
 \label{fig:sp_entropy_2d_unscaled}
\end{figure}

The negative Casimir entropy for the sphere-plane geometry with perfect metals
is shown in Fig.~\ref{fig:sp_entropy_2d_unscaled} as a function of the ratio of
the distance $L$ between the sphere's surface and the plane to the radius $R$
of the sphere, and of the temperature $T$. The Casimir entropy vanishes along
the dashed line. Below this line, the Casimir entropy is negative while it is
positive above. The grey area indicates parameter regions where the Casimir
entropy has not been evaluated.

According to Fig.~\ref{fig:sp_entropy_2d_unscaled}, the Casimir entropy becomes
minimal at $L/R\approx0.27$ and $2\pi\mathcal{L}k_BT/\hbar c\approx0.93$. At a
fixed temperature, the Casimir entropy becomes positive if the radius $R$ is
sufficiently large. In contrast, for small radii, the Casimir entropy will
always be negative for temperatures below a threshold value to be specified in
the discussion of the free energy (\ref{eq:freeEnergySpherePlane}), see below.

\begin{figure}
 \begin{center}
  \includegraphics[width=\columnwidth]{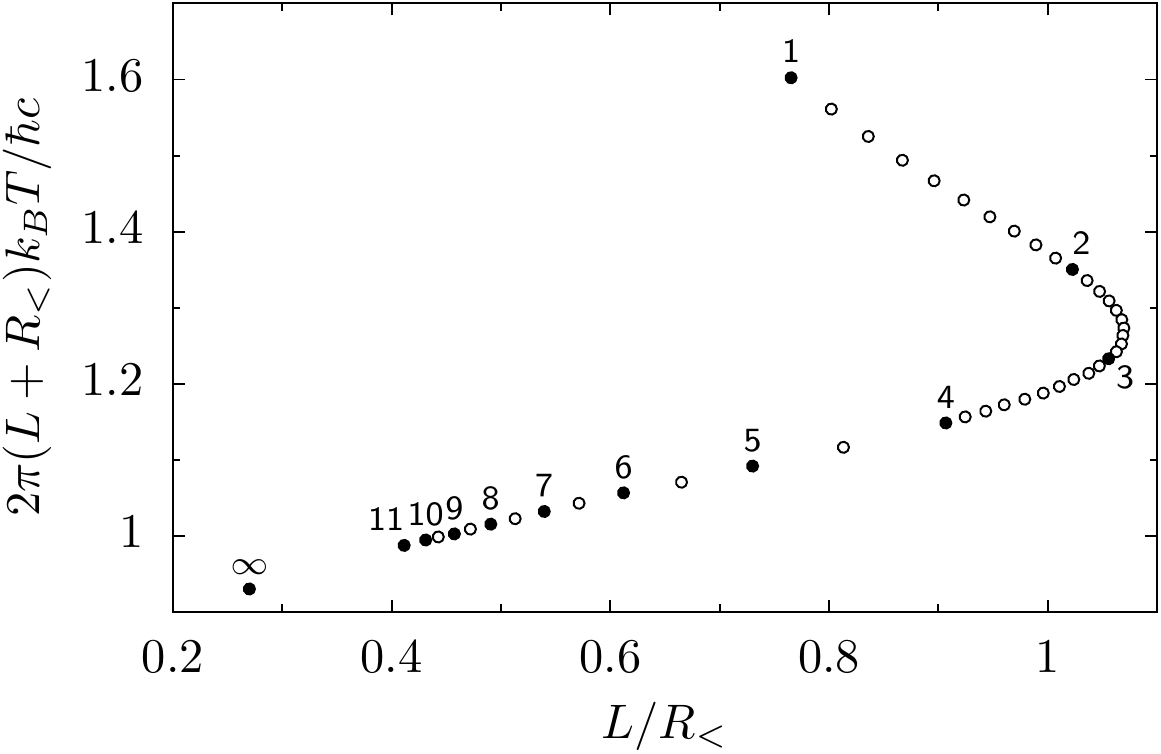}
 \end{center}
 \caption{The transition from perfect-metal sphere-sphere to sphere-plane
  configurations is illustrated by the position of the minimum of the entropy
  as a function of the surface-to-surface distance $L$ and the temperature $T$.
  The points refer to different ratios of the radii $R_>$ and $R_<$ of the larger
  and smaller sphere, respectively. For the filled circles, the ratio of radii
  is indicated in the plot. The point marked by $\infty$ corresponds to the
  position of the minimum of the Casimir entropy in
  Fig.~\ref{fig:sp_entropy_2d_unscaled}.}
 \label{fig:ss2sp_extremum}
\end{figure}

It would be expected that the sphere-plane configuration can be obtained from
the sphere-sphere configuration by letting the radius of one of the two spheres
go to infinity. To illustrate this transition, we show in
Fig.~\ref{fig:ss2sp_extremum} the position of the minimum of the Casimir
entropy for perfect metals as a function of the geometric parameters and the
temperature for two spheres with radii $R_<$ and $R_>$ for the smaller and
larger sphere, respectively. For the filled circles, the corresponding values
of the ratio $R_>/R_<$ are indicated, ranging from $R_>/R_<=1$ for spheres of
equal radius to the extreme case $R_>/R_<=\infty$ for which the position of the
entropy minimum for the plane-sphere geometry is shown. The data clearly
indicate a smooth transition between the sphere-sphere and the plane-sphere
configuration.

\begin{figure}
 \begin{center}
  \includegraphics[width=\columnwidth]{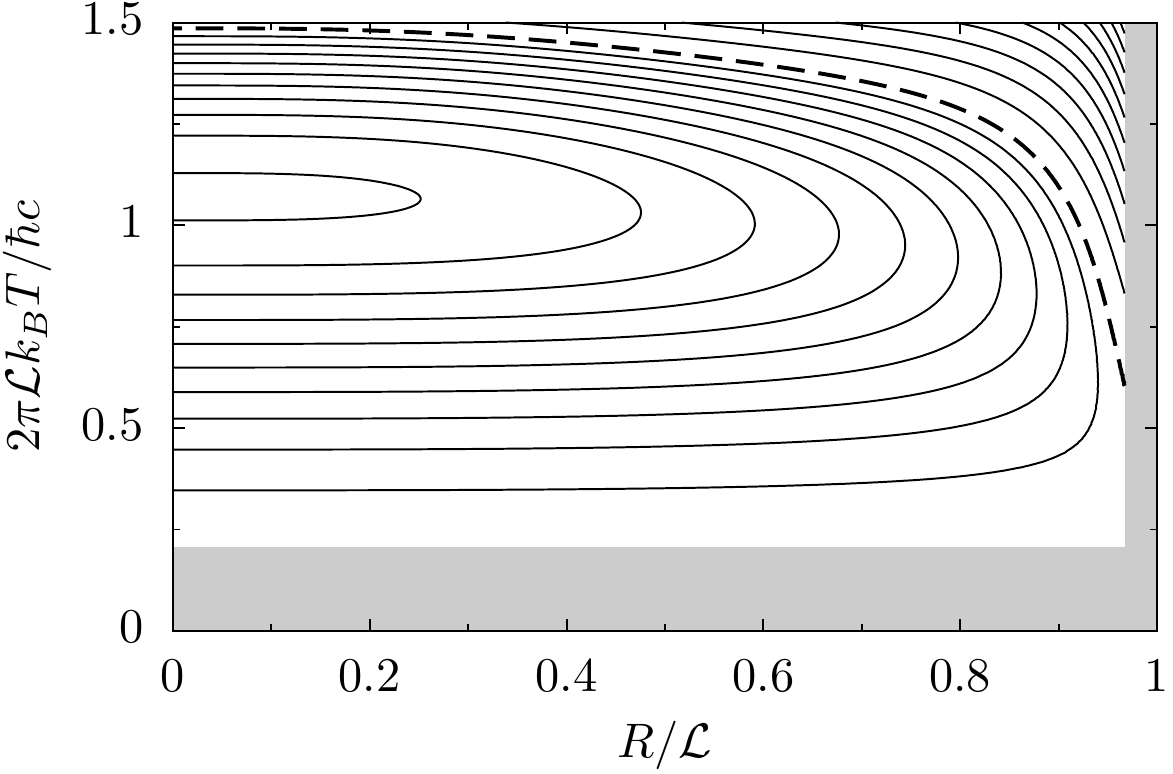}
 \end{center}
 \caption{The Casimir entropy in the sphere-plane geometry multiplied by a 
  factor $(\mathcal{L}/R)^3$ is displayed for perfect metals as a function
  of the inverse of the distance $\mathcal{L}$ between sphere's center
  and the plane and the temperature $T$. The entropy vanishes along the
  dashed line. Below this line, the entropy is negative and changes in
  steps of 0.00125. Above the dashed line, the entropy is positive and
  changes in steps of 0.0025. No data have been calculated in the grey
  region.}
 \label{fig:sp_entropy_2d_scaled}
\end{figure}

As far as the physical origin of the negative Casimir entropy is concerned, the
presentation of the data in Fig.~\ref{fig:sp_entropy_2d_unscaled} is somewhat
misleading because for small sphere radius, the entropy decreases with the
volume of the sphere. It is thus appropriate to scale the entropy with
$(\mathcal{L}/R)^3$. The result is depicted in
Fig.~\ref{fig:sp_entropy_2d_scaled} where the dashed line again indicates a
vanishing Casimir entropy and negative values of the Casimir entropy are found
below the dashed line.  Note that in this plot, in contrast to
Fig.~\ref{fig:sp_entropy_2d_unscaled}, small sphere radii are on the left side.
The minimum of the scaled Casimir entropy lies at $R=0$. We can thus conclude
that the large-distance limit $L, \mathcal{L}\gg R$ is well suited for an
analysis of the negative Casimir entropy.

By means of the thermodynamic relation
\begin{equation}
\label{eq:defEntropy}
\mathcal{S} = -\frac{\partial\mathcal{F}}{\partial T}\,,
\end{equation}
the Casimir entropy $\mathcal{S}$ for the sphere-plane configuration with
perfect metals in the large-distance limit can easily be obtained from the
expression for the Casimir free energy $\mathcal{F}$ in this limit
\cite{Canaguier2010}
\begin{equation}
\label{eq:freeEnergySpherePlane}
\mathcal{F} = \frac{3\hbar c}{16\pi}\frac{R^3}{\mathcal{L}^4}
	\left[g(\nu)\cosh(\nu)+g(\nu)^2+g(\nu)^3\cosh(\nu)\right]\,,
\end{equation}
where we introduced the abbreviation
\begin{equation}
\label{eq:def_g}
g(\nu)=\frac{\nu}{\sinh(\nu)}\,.
\end{equation}
Taking the derivative with respect to temperature, one finds negative values
for the Casimir entropy for temperatures satisfying $0<\nu\lessapprox1.486$ in
agreement with the data shown in Fig.~\ref{fig:sp_entropy_2d_scaled} for small
values of $R$. In order to obtain information about the physical origin of the
negative Casimir entropy, in Sect.~\ref{subsec:sp_large_distance} we will
decompose the free energy (\ref{eq:freeEnergySpherePlane}) and the entropy
derived from it into contributions arising from different scattering channels.

\begin{figure}
 \begin{center}
  \includegraphics[width=\columnwidth]{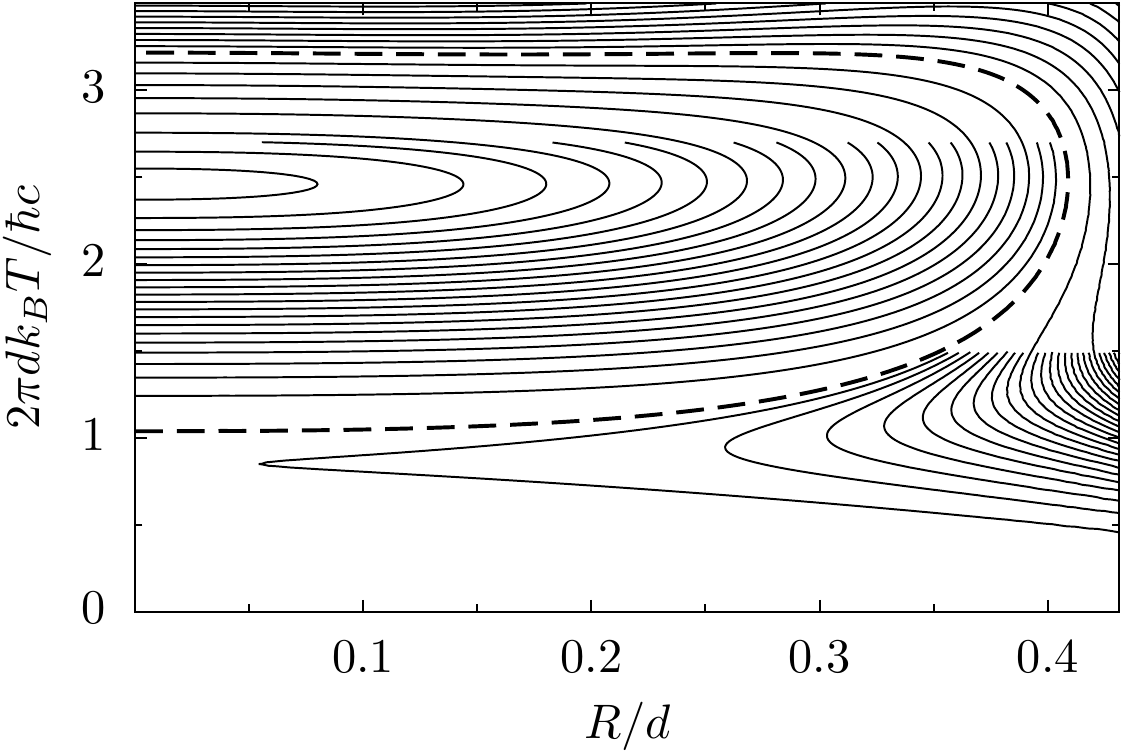}
 \end{center}
 \caption{The Casimir entropy in the sphere-sphere geometry multiplied with
   $(d/R)^6$ is depicted for perfect metals as a function of the inverse of
   the distance $d$ between the centers of the spheres and of the temperature
   $T$ for spheres of equal radii, $R_1=R_2=R$. The entropy vanishes along the
   dashed line. Inside the region bounded by the dashed line, the entropy is
   negative and changes in steps of 0.025. Outside of the dashed line, the
   entropy is positive and changes in steps of 0.005. The density of the contour
   lines has been decreased at higher temperatures to improve the clarity of
   the plot.}
 \label{fig:ss_entropy_2d}
\end{figure}

For the sphere-sphere geometry, the region of negative Casimir entropy for
perfect-metal spheres of equal radii is displayed in
Fig.~\ref{fig:ss_entropy_2d} where the entropy is scaled by $(d/R)^6$.  The
dashed line separates the regions of negative and positive Casimir entropies
with negative values appearing inside the region delimited by the dashed line.
As an obvious difference to the sphere-plane geometry shown in
Fig.~\ref{fig:sp_entropy_2d_scaled} we observe that while in the latter case
negative values of the Casimir entropy are found down to the lowest
temperatures, this is not the case in the sphere-sphere geometry. This fact has
already been noted in \cite{Rodriguez2011}. Furthermore, the region of negative
Casimir entropies ends before $R/d$ reaches its maximum value of $1/2$.

Despite these differences, Fig.~\ref{fig:ss_entropy_2d} suggests that again
interesting insights into the physics of the negative Casimir entropy can be
obtained from the large-distance limit $d\gg R$. The free energy for
perfect-metal spheres in this limit is given by \cite{Rodriguez2011}
\begin{equation}
\label{eq:freeEnergySphereSphere}
\begin{aligned}
\mathcal{F} &= -\frac{\hbar c}{16\pi}\frac{R^6}{d^7}\left[30g(\nu)\cosh(\nu)
                  +30g(\nu)^2\right.\\
&\qquad\qquad+29g(\nu)^3\cosh(\nu)+9g(\nu)^4\left(2\cosh(\nu)^2+1\right)\\\,.
&\qquad\qquad\left.+9g(\nu)^5\cosh(\nu)\left(\cosh(\nu)^2+2\right)\right]\,,
\end{aligned}
\end{equation}
where $g(\nu)$ was defined in (\ref{eq:def_g}). A decomposition of this result
and the corresponding entropy in terms of the scattering channels will be
carried out in Sect.~\ref{subsec:ss_large_distance}.

\section{Large-distance approximation}
\label{sec:largeDistanceApproximation}

Within the scattering approach, the Casimir free energy can be expressed as
\begin{equation}
\label{eq:matsubaraFreeEnergy}
\mathcal{F} = 2k_BT\sum_{n=0}^\infty{}'\sum_{m=0}^\infty{}'\ln\left(\det\left[
1-\mathcal{M}^{(m)}(\xi_n)\right]\right)\,.
\end{equation}
The first sum runs over the Matsubara frequencies $\xi_n=2\pi n/\hbar\beta$
with $\beta=1/k_BT$. The prime indicates that only one half of the
$n=0$-term should be taken. The geometries depicted in Fig.~\ref{fig:geometry}
are symmetric under rotations around the $z$-axis, thus allowing us to
decompose the scattering problem into subspaces of fixed eigenvalues $m$ of the
$z$-component of the angular momentum. Since the sign of $m$ is irrelevant
for the scattering process, we sum only over positive values of $m$, leading
to a prefactor 2 except for $m=0$ as again indicated by a prime.

The matrix $\mathcal{M}^{(m)}(\xi_n)$ describes a roundtrip scattering process
between the two scattering objects at imaginary frequency $i\xi_n$ in the
subspace of the $z$-component of the angular momentum characterized by $m$.
The roundtrip scattering operator
\begin{equation}
\label{eq:roundtripMatrixelement}
\mathcal{M}^{(m)}= \mathcal{R}_1^{(m)}\mathcal{T}_{12}^{(m)}\mathcal{R}_2^{(m)}
                   \mathcal{T}_{21}^{(m)}
\end{equation}
contains four building blocks, namely the translation operator
$\mathcal{T}_{21}$ from the reference frame of object 1 to that of object 2,
the reflection operator $\mathcal{R}_2$ on object 2, the reverse translation
operator $\mathcal{T}_{12}$, and finally the reflection operator
$\mathcal{R}_1$ on object 1. While keeping the quantum number $m$ unchanged,
these operators will in general modify the other parameters of the scattered
modes like their polarization and, for spherical waves, their angular momentum
quantum number $\ell$ or, for plane waves, their wave vector $\mathbf{k}$.

We argued in the previous section that the limit of large distances between
the scattering objects is appropriate to analyze the origin of negative Casimir
entropies. To be specific, we assume that the distance $D$ is much larger than
the sphere radius $R$. In the case of two spheres, $R$ refers to the larger of
the two radii. As we shall see now, this limit allows us to quantify the
contribution of each scattering channel to the Casimir entropy.

Since the matrix elements of the translation operators for imaginary
frequencies decay exponentially with $D$, the highest relevant frequencies are
of the order of $c/D$. The reflection on a sphere will then contribute a factor
$(R/D)^{2\ell+1}$ where $\ell$ denotes the order of the multipole wave
scattered at the sphere. As a consequence, we may restrict the scattering at a
sphere to $\ell=1$. In the large-distance limit, the Casimir entropy for the
sphere-plane geometry and the sphere-sphere geometry thus scales with
$(R/\mathcal{L})^3$ and $(R/d)^6$, respectively. These factors are at the
origin of the scaling employed in the previous section.

Furthermore, in the large-distance limit the matrix elements of the roundtrip
operator are very small and we may expand the logarithm in
(\ref{eq:matsubaraFreeEnergy}). In view of
$\text{tr}(\mathsf{A})=\log[\det(\exp(\mathsf{A}))]$ with $\text{tr}$ denoting
the trace, we then obtain
\begin{equation}
\label{eq:freeEnergySum2}
\mathcal{F} = -k_BT\sum_{n=0}^\infty{}'\sum_{P,P'}\left[
\mathcal{M}^{(0)}_{P,P}(\xi_n)\delta_{P,P'}
+2\mathcal{M}^{(1)}_{P,P'}(\xi_n)\right]\,,
\end{equation}
where $P$ and $P'$ denote the polarizations on the two scattering objects.
Even though we restrict our considerations to $\ell=1$ on each sphere, the
translation operators $\mathcal{T}$ present in the matrix elements
$\mathcal{M}^{(m)}_{P,P'}$ implicitly give rise to a sum over other multipole
moments $\ell'$ or, in the sphere-plane geometry, to an integral over the
projection $\mathbf{k}_\parallel$ of the wave vector onto the plane.

The polarizations $P$ and $P'$ in (\ref{eq:freeEnergySum2}) can be either
transverse electric (TE) or transverse magnetic (TM) and correspond to the mode
polarizations on the two scatterers. Note that on a sphere, the TE and TM
polarizations are sometimes referred to as H and E polarizations, respectively.
While in general, the polarization may change in the course of the scattering
process, this is not the case for $m=0$.  In this case, for a TE (TM) mode on
the sphere the electric (magnetic) field will only have a component in the
plane perpendicular to the $z$ direction. As a consequence, the expansion of
this mode on the other sphere or the plane will contain exclusively TE (TM)
modes and no polarization change ensues.

We thus have to account for three essentially different scattering channels.
The channels where the polarization is conserved contribute for $m=0$ and
$m=1$. A third channel involves a change of polarization and is restricted to
$m=1$. The latter channel is of particular interest for our discussion for
the following reason.

Focussing on the contribution of the translation operator, it is clear that in
the absence of a shift, the polarization of the mode cannot change. In
(\ref{eq:freeEnergySum2}), for dimensional reasons, the shift can only appear
in the combination $\xi_nD/c$. As a consequence, the $n=0$ term will vanish if
the polarizations $P$ and $P'$ differ, implying in turn that the free energy
contribution of this channel will vanish at high temperatures. Since the
contribution to the Casimir free energy of the polarization-changing channel at
zero temperature is negative and turns out to be monotonically increasing, its
contribution to the Casimir entropy in view of (\ref{eq:defEntropy}) will
always be negative.

While the differences between the scattering channels discussed so far were
mainly due to the translation operators $\mathcal{T}$, the reflection
properties of the sphere(s) offer an interesting way to select channels. The
reflection at a perfectly conducting (PC) sphere of radius $R$ to leading order
in $\xi R/c$ is dominated by the Mie coefficients with $\ell=1$. For the TM
mode, the Mie coefficient is then given by
\begin{equation}
\label{eq:mie_a1}
a_1^\text{PC}(\xi) = -\frac{2}{3}\left(\frac{\xi R}{c}\right)^3 + O(\xi^5)
\end{equation}
and for the TE mode it reads
\begin{equation}
\label{eq:mie_b1}
b_1^\text{PC}(\xi) = \frac{1}{3}\left(\frac{\xi R}{c}\right)^3 + O(\xi^5)\,.
\end{equation}
We thus have $a_1^{PC}=-2b_1^{PC}$, leading to simple numerical relations
between the contributions of the polarization-conserving scattering channels.

In contrast, for spheres made of a metal described by the Drude model (D),
one finds
\begin{equation}
a_1^\text{D}(\xi) = -\frac{2}{3}\left(\frac{\xi R}{c}\right)^3 + O(\xi^4)\,,
\end{equation}
while $b_1^\text{D}(\xi)$ is of order $(\xi R/c)^4$ and therefore negligible
within the large-distance approximation. Switching from a perfectly conducting
sphere to a Drude metal sphere allows us to suppress the reflection of TE modes 
on the sphere. In particular, the polarization-changing channel will become
irrelevant as we will explain in more detail in Sect.~\ref{sec:drude}.

These general considerations and their consequences for the Casimir entropy
will be worked out more explicitly in the following section dealing
with the specific geometries shown in Fig.~\ref{fig:geometry}. 

\section{Casimir free energy and entropy for perfect conductors}
\subsection{Sphere-plane geometry}
\label{subsec:sp_large_distance}

Within the large-distance approximation, we only need the matrix elements of
the roundtrip operator for $m=0$ and $1$ to obtain the free energy by means of
(\ref{eq:freeEnergySum2}). We then can make use of (\ref{eq:defEntropy}) to
obtain the entropy. For the sphere-plane geometry, the matrix element of the
roundtrip operator can easily be obtained from the expressions given in
Ref.~\cite{Canaguier2010}. For $m=0$, we obtain
\begin{equation}
\label{eq:m0tmtm}
\mathcal{M}_{TM,TM}^{(0)} = \frac{1}{4}\left(\frac{R}{\mathcal{L}}\right)^3
(1+2\tilde\xi)\exp(-2\tilde\xi)
\end{equation}
and
\begin{equation}
\label{eq:m0tete}
\mathcal{M}_{TE,TE}^{(0)} = \frac{1}{2}\mathcal{M}_{TM,TM}^{(0)}
\end{equation}
with the dimensionless frequency $\tilde\xi$ introduced in
(\ref{eq:dimensionlessFrequency}).

For $m=1$, the matrix elements for the roundtrips where polarization is
conserved are
\begin{equation}
\label{eq:m1tmtm}
\mathcal{M}_{TM,TM}^{(1)} = \frac{1}{8}\left(\frac{R}{\mathcal{L}}\right)^3
(1+2\tilde\xi+2\tilde\xi^2)\exp(-2\tilde\xi)
\end{equation}
and
\begin{equation}
\label{eq:m1tete}
\mathcal{M}_{TE,TE}^{(1)} = \frac{1}{2}\mathcal{M}_{TM,TM}^{(1)}\,.
\end{equation}
For roundtrips involving a change of polarization, one finds
\begin{equation}
\label{eq:m1tmte}
\mathcal{M}_{TM,TE}^{(1)} = \frac{1}{4}\left(\frac{R}{\mathcal{L}}\right)^3
\tilde\xi^2\exp(-2\tilde\xi)
\end{equation}
and
\begin{equation}
\label{eq:m1tetm}
\mathcal{M}_{TE,TM}^{(1)} = \frac{1}{2}\mathcal{M}_{TM,TE}^{(1)}\,.
\end{equation}
Here, the first subscript of the roundtrip operator $\mathcal{M}$ refers to the
polarization on the sphere while the second subscript indicates the
polarization on the plane. The matrix elements (\ref{eq:m1tmte}) and
(\ref{eq:m1tetm}) for scattering processes involving a change of polarization
vanish in the limit of vanishing frequency $\tilde\xi$ as discussed in the
previous section.

Making use of (\ref{eq:freeEnergySum2}), we can decompose the Casimir free
energy for perfect metal sphere and plane into contributions from the different
scattering channels according to
\begin{equation}
\label{eq:PSfreeEnergyGeneral}
\begin{aligned}
\mathcal{F} &= -\frac{\hbar c}{2\pi\mathcal{L}}
               \left(\frac{R}{\mathcal{L}}\right)^3
               \left[f^{(0)}_{TM,TM}+f^{(0)}_{TE,TE}\right.\\
&\qquad\left.+f^{(1)}_{TM,TM}+f^{(1)}_{TE,TE}
             +f^{(1)}_{TM,TE}+f^{(1)}_{TE,TM}\right]\,.
\end{aligned}
\end{equation}
Evaluating the corresponding Matsubara sums, we find for $m=0$ from
(\ref{eq:m0tmtm})
\begin{equation}
\label{eq:sp_f0tmtm}
f^{(0)}_{TM,TM} = \frac{1}{8}\left[g(\nu)\cosh(\nu)+g(\nu)^2\right]\,.
\end{equation}
Here, we have made use of the dimensionless temperature
(\ref{eq:dimensionlessTemperature}) and the function $g$ defined in (\ref{eq:def_g}).
For $m=1$, we obtain from (\ref{eq:m1tmtm}) and (\ref{eq:m1tmte})
\begin{equation}
\label{eq:sp_f1tmtm}
f^{(1)}_{TM,TM} = \frac{1}{8}\left[g(\nu)\cosh(\nu)+g(\nu)^2+g(\nu)^3\cosh(\nu)
                  \right]
\end{equation}
and
\begin{equation}
\label{eq:sp_f1tmte}
f^{(1)}_{TM,TE} = \frac{1}{8}g(\nu)^3\cosh(\nu)\,,
\end{equation}
respectively. The remaining three contributions in
(\ref{eq:PSfreeEnergyGeneral}) are related to the expressions just given by a
factor of 1/2 according to the relations (\ref{eq:m0tete}), (\ref{eq:m1tete}),
and (\ref{eq:m1tetm}). Summing up all terms, we recover the free energy
(\ref{eq:freeEnergySpherePlane}) for sphere and plane made of perfect
conductors obtained earlier in Ref.~\cite{Canaguier2010}. The decomposition
in terms of the different scattering channels is found to be in agreement with
the expressions obtained in Ref.~\cite{Milton2014}. In order to make the
connection, one sets the electric and magnetic polarizability equal to $R^3$ and 
$-R^3/2$, respectively. For vanishing polarizability in the transverse direction,
one finds the contribution for $m=0$ while an isotropic polarizability yields
the sum of the contributions from $m=0$ and $1$.

It is now straightforward to obtain the contributions of the various scattering
channels to the Casimir entropy. Expressing the Casimir entropy in terms of a
rescaled Casimir entropy $s$ according to
\begin{equation}
S = k_B\left(\frac{R}{\mathcal{L}}\right)^3s\,,
\end{equation}
the definition of the entropy (\ref{eq:defEntropy}) turns into
\begin{equation}
s=\frac{\partial f}{\partial\nu}\,.
\end{equation}
We then obtain from (\ref{eq:sp_f0tmtm}), (\ref{eq:sp_f1tmtm}), and
(\ref{eq:sp_f1tmte}) 
\begin{equation}
\label{eq:sp_s0tmtm}
s^{(0)}_{TM,TM} = \frac{1}{8\nu}\left[g(\nu)\cosh(\nu)+g(\nu)^2
                  -2g(\nu)^3\cosh(\nu)\right]
\end{equation}
\begin{equation}
\label{eq:sp_s1tmtm}
\begin{aligned}
s^{(1)}_{TM,TM} &= \frac{1}{8\nu}\left[g(\nu)\cosh(\nu)+g(\nu)^2
                  +g(\nu)^3\cosh(\nu)\right.\\
		  &\qquad\qquad\left.-g(\nu)^4\big(2\cosh^2(\nu)+1\big)\right]
\end{aligned}
\end{equation}
\begin{equation}
\label{eq:sp_s1tmte}
s^{(1)}_{TM,TE} = \frac{1}{8\nu}\left[3g(\nu)^3\cosh(\nu)
                  -g(\nu)^4\big(2\cosh^2(\nu)+1\big)\right]\,.
\end{equation}
The remaining three contributions to the Casimir entropy can be obtained by
simple multiplication with a factor $1/2$ as before for the free energy and the
matrix elements of the roundtrip operator.

\begin{figure}
 \begin{center}
  \includegraphics[width=\columnwidth]{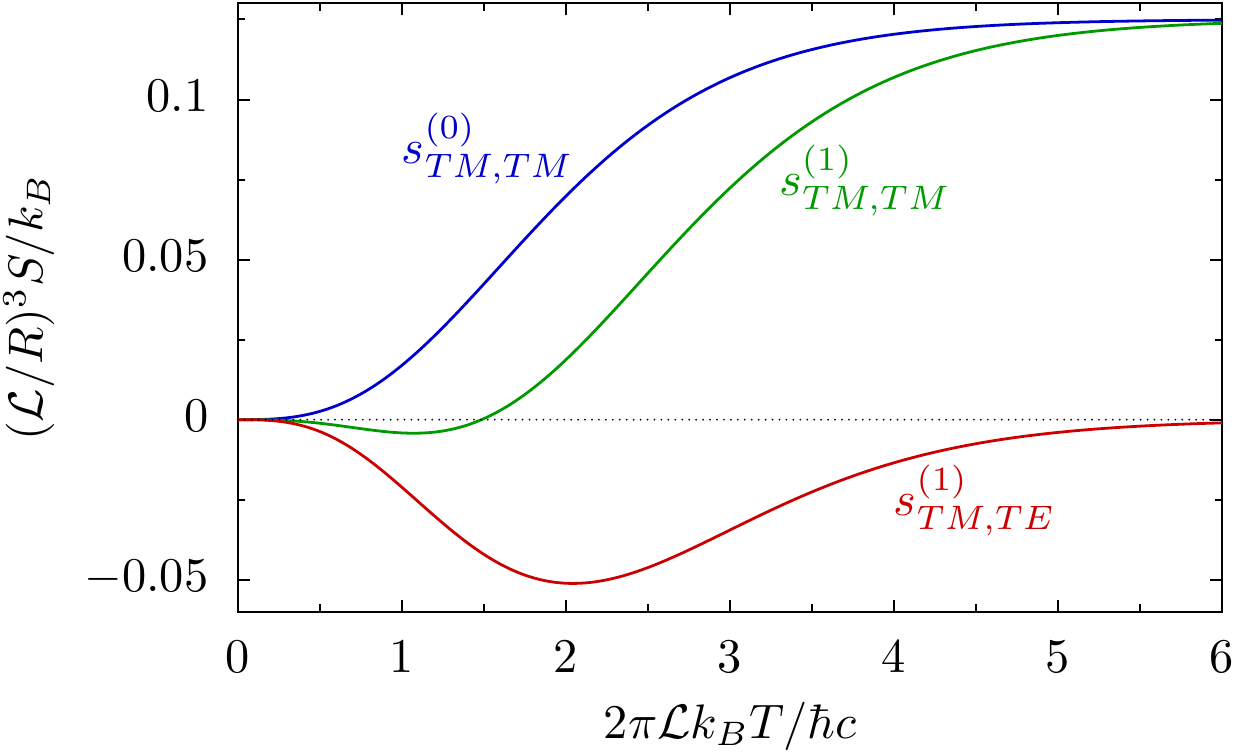}
 \end{center}
 \caption{The contributions (\ref{eq:sp_s0tmtm}), (\ref{eq:sp_s1tmtm}), and
  (\ref{eq:sp_s1tmte}) to the Casimir entropy in the plane-sphere geometry
  arising from the scattering channels with TM polarization on the sphere are
  shown as a function of the dimensionless temperature. While $S^{(0)}_{TM,TM}$ 
  is always positive and $S^{(1)}_{TM,TM}$ becomes only slightly negative for
  small temperatures, $S^{(1)}_{TM,TE}$ is negative for all temperatures.}
 \label{fig:sp_entropy_largedist}
\end{figure}

In Fig.~\ref{fig:sp_entropy_largedist}, the temperature dependence of the
contributions (\ref{eq:sp_s0tmtm}), (\ref{eq:sp_s1tmtm}), and
(\ref{eq:sp_s1tmte}) from the channels with TM polarization on the sphere to
the Casimir entropy are shown. Among the polarization-conserving channels, the
$m=0$ contribution is positive for all temperatures. The $m=1$ contribution,
while being slightly negative at sufficiently small temperatures, in
combination with the $m=0$ contribution will still always lead to positive
values of the entropy. In order to arrive at a negative entropy, one needs the
polarization-changing mode. In fact, $S^{(1)}_{TM,TE}$ is negative for all
temperatures as was already conjectured in
Sect.~\ref{sec:largeDistanceApproximation}.

The negative contribution of the polarization-changing channel is indeed
sufficiently large to render the sum of all contributions negative. This
can clearly be seen from the low-temperature expansions of the entropies
(\ref{eq:sp_s0tmtm})--(\ref{eq:sp_s1tmte})
\begin{align}
s^{(0)}_{TM,TM} &= \frac{1}{45}\nu^3-\frac{2}{315}\nu^5+O(\nu^7)\\
s^{(1)}_{TM,TM} &= -\frac{1}{90}\nu^3+\frac{1}{105}\nu^5+O(\nu^7)\\
s^{(1)}_{TM,TE} &= -\frac{1}{30}\nu^3+\frac{1}{63}\nu^5+O(\nu^7)\,.
\end{align}

Taking all six channels between perfect conductor plane and sphere
into account, we obtain for the low-temperature expansion of the Casimir
entropy
\begin{equation}
\label{eq:sp_s_lowtemp}
s = -\frac{1}{30}\nu^3+\frac{1}{35}\nu^5+O(\nu^7)\,.
\end{equation}
In the large-distance limit, we thus obtain negative values for the
Casimir entropy at sufficiently low temperatures. In the limit of vanishing
temperature, the Casimir entropy goes to zero in agreement with the third law
of thermodynamics.

\subsection{Sphere-sphere geometry}
\label{subsec:ss_large_distance}

We now turn to the discussion of the Casimir free energy and entropy of the
sphere-sphere configuration depicted in Fig.~\ref{fig:geometry}b. In order to
determine the Casimir free energy in the limit of large distance,
$d\gg R_1, R_2$, from the expression (\ref{eq:freeEnergySum2}), we first need
to determine the matrix elements $\mathcal{M}^{(m)}_{P,P'}$ of the round-trip
operator with $m=0$ and $1$.

As discussed in Sect.~\ref{sec:largeDistanceApproximation}, the large-distance
approximation implies that on the spheres, we can restrict the field modes to
dipole spherical waves, $\ell=1$. For the spheres, the matrix elements of the
reflection operators are thus simply given by the Mie coefficients
(\ref{eq:mie_a1}) and (\ref{eq:mie_b1}), and we have
\begin{equation}
\label{eq:mmtmtm}
\mathcal{R}^{(m)}_{TM,TM} = -\frac{2}{3}\left(\frac{R}{d}\right)^3\tilde\xi^3
\end{equation}
and
\begin{equation}
\label{eq:mmtete}
\mathcal{R}^{(m)}_{TE,TE} = \frac{1}{3}\left(\frac{R}{d}\right)^3\tilde\xi^3\,,
\end{equation}
where $R$ is the sphere radius and $\tilde\xi$ is the dimensionless imaginary
frequency (\ref{eq:dimensionlessFrequency}). In contrast to the reflection on a
plane in the preceding subsection, the polarization of a spherical wave remains
unchanged by the reflection at a sphere.

However, the translation operator for spherical waves with wave vector $k$ from
the center of one sphere to the center of the other sphere leads to a mixing of
polarizations. The general expressions for the matrix elements of the
translation operator \cite{Cruzan1962} can be simplified if the translation is
performed along the $z$-axis \cite{Bruning1969,Bruning1971} as is the case in
the setup displayed in Fig.~\ref{fig:geometry}b. Within the large-distance
approximation, we find for the channels conserving the polarization
\begin{equation}
\label{eq:tmpp}
\begin{aligned}
\mathcal{T}^{(m)}_{P,P} &= (-1)^m\frac{3}{4}\sum_{\ell'=0, 2}i^{-\ell'}
[4-\ell'(\ell'+1)](2\ell'+1)\\
&\qquad\times\begin{pmatrix} 1 & 1 & \ell'\\ 0 & 0 & 0 \end{pmatrix}
            \begin{pmatrix} 1 & 1 & \ell'\\ m & -m & 0 \end{pmatrix}
h_{\ell'}^{(1)}(kd)\,.
\end{aligned}
\end{equation}
For a change of polarization, $P\neq P'$, the matrix element vanishes for
$m=0$ in agreement with the argument given in
Sect.~\ref{sec:largeDistanceApproximation}, while for $m=1$ we have
\begin{equation}
\label{eq:tmppp}
\begin{aligned}
\mathcal{T}^{(1)}_{P,P'} &= \pm\frac{3}{2}ikd\sum_{\ell'=0, 2}i^{-\ell'}
(2\ell'+1)\\
&\qquad\times\begin{pmatrix} 1 & 1 & \ell'\\ 0 & 0 & 0 \end{pmatrix}
            \begin{pmatrix} 1 & 1 & \ell'\\ 1 & -1 & 0 \end{pmatrix}
h_{\ell'}^{(1)}(kd)\,.
\end{aligned}
\end{equation}
The first two factors in the second lines of (\ref{eq:tmpp}) and
(\ref{eq:tmppp}) denote Wigner $3j$ symbols while $h_\ell^{(1)}$ is a spherical
Bessel function of the third kind. The overall sign in (\ref{eq:tmppp}) is
positive or negative depending on whether the translation is performed in
positive or negative $z$-direction, respectively.

Employing the relation
\begin{equation}
h_\ell^{(1)}(ix) = -\frac{2}{\pi}i^{-\ell}k_\ell(x)
\end{equation}
between the spherical Bessel function of the third kind and the modified
spherical Bessel function $k_\ell$, it is straightforward to express the
matrix elements of the translation operator in imaginary frequency $\xi$
needed in order to evaluate the Matsubara sum (\ref{eq:freeEnergySum2}).
With the explicit expressions for the modified spherical Bessel functions
\begin{equation}
\begin{aligned}
k_0(x) &= \frac{\pi}{2}\frac{\exp(-x)}{x}\\
k_2(x) &= \frac{\pi}{2}\left(\frac{1}{x}+\frac{3}{x^2}+\frac{3}{x^3}\right)
          \exp(-x)
\end{aligned}
\end{equation}
the matrix elements needed in the following are then found to read
\begin{equation}
\mathcal{T}^{(0)}_{P,P} = 3\left(\frac{1}{\tilde\xi^3}
  +\frac{1}{\tilde\xi^2}\right)\exp(-\tilde\xi)
\end{equation}
and
\begin{equation}
\mathcal{T}^{(1)}_{P,P} = -\frac{3}{2}\left(\frac{1}{\tilde\xi^3}
  +\frac{1}{\tilde\xi^2}+\frac{1}{\tilde\xi}\right)\exp(-\tilde\xi)
\end{equation}
for $P=TE, TM$ while for differing polarizations $P$ and $P'$ one finds
\begin{equation}
\label{eq:t1ppp}
\mathcal{T}^{(1)}_{P,P'} = \pm\frac{3}{2}\left(\frac{1}{\tilde\xi^2}
+\frac{1}{\tilde\xi}\right)\exp(-\tilde\xi)\,.
\end{equation}
In the last equation, the sign depends on the direction of translation with
respect to the $z$-axis as discussed in the context of (\ref{eq:tmppp}).

We note that even though these matrix elements diverge for vanishing frequency
$\tilde\xi$, their products with the matrix elements (\ref{eq:mmtmtm}) and
(\ref{eq:mmtete}) of the reflection operator remain finite. While the
combination of the two matrix elements yields a nonzero value if the
polarization is conserved, the product of the matrix element (\ref{eq:t1ppp})
for changing polarization with one of the reflection matrix elements
(\ref{eq:mmtmtm}) and (\ref{eq:mmtete}) goes to zero for vanishing frequency 
$\tilde\xi$. 

As a consequence, the contribution of the polarization-changing channels to the
Casimir free energy and entropy vanishes at high temperatures where the
Matsubara sum (\ref{eq:freeEnergySum2}) is dominated by the $n=0$ term. Already
at this point, we can therefore expect the same qualitative temperature
dependence of the contributions of the polarization-changing channels as in the
plane-sphere geometry. These channels will thus play the same crucial role for
an overall negative Casimir entropy also for the sphere-sphere geometry.  The
difference between the polarization-conserving and polarization-changing
channels can be traced back to an extra factor $kd$ appearing in the front of
the right-hand side of (\ref{eq:tmppp}) compared to (\ref{eq:tmpp}). This
factor ensures that for vanishing translation, $d=0$, no change of polarization
occurs as was argued in Sect.~\ref{sec:largeDistanceApproximation}.

With the matrix elements listed above, it is straightforward to evaluate
the contributions of the various channels to the Casimir free energy by means of
(\ref{eq:roundtripMatrixelement}) and (\ref{eq:freeEnergySum2}). We decompose
the Casimir free energy into the contributions from the various scattering
channels
\begin{equation}
\label{eq:SSfreeEnergyGeneral}
\begin{aligned}
\mathcal{F} &= -\frac{\hbar c}{2\pi d}
               \left(\frac{R_1R_2}{d^2}\right)^3
               \left[f^{(0)}_{TM,TM}+f^{(0)}_{TE,TE}\right.\\
&\qquad\left.+f^{(1)}_{TM,TM}+f^{(1)}_{TE,TE}
             +f^{(1)}_{TM,TE}+f^{(1)}_{TE,TM}\right]\,.
\end{aligned}
\end{equation}
The contributions to (\ref{eq:SSfreeEnergyGeneral}) arising from
polarization-conserving channels are given by
\begin{equation}
\label{eq:ss_f0tmtm}
f^{(0)}_{TM,TM} = 2g(\nu)\cosh(\nu)+2g(\nu)^2+g(\nu)^3\cosh(\nu)
\end{equation}
and
\begin{equation}
\label{eq:ss_f1tmtm}
\begin{aligned}
f^{(1)}_{TM,TM} &= \frac{1}{2}\left[2g(\nu)\cosh(\nu)+2g(\nu)^2
                   +3g(\nu)^3\cosh(\nu)\right.\\
&\qquad\quad+g(\nu)^4\big(2\cosh^2(\nu)+1\big)\\
&\qquad\quad\left.+g(\nu)^5\cosh(\nu)\big(\cosh^2(\nu)+2\big)\right]
\end{aligned}
\end{equation}
together with
\begin{equation}
f^{(m)}_{TE,TE} = \frac{1}{4}f^{(m)}_{TM,TM}\,.
\end{equation}
The contributions of the polarization-changing channels are
\begin{equation}
\begin{aligned}
f^{(1)}_{TM,TE} &= \frac{1}{4}\left[g(\nu)^3\cosh(\nu)
                   +g(\nu)^4\big(2\cosh^2(\nu)+1)\right.\\
&\qquad\quad\left.+g(\nu)^5\cosh(\nu)\big(\cosh^2(\nu)+2\big)\right]
\end{aligned}
\end{equation}
and
\begin{equation}
f^{(1)}_{TE,TM} = f^{(1)}_{TM,TE}\,.
\end{equation}
In these results, we make use of the dimensionless temperature
(\ref{eq:dimensionlessTemperature}) and the function $g$ defined in
(\ref{eq:def_g}).  The total Casimir free energy obtained from these
expressions agrees with the result (\ref{eq:freeEnergySphereSphere}) and
was already given in Ref.~\cite{Rodriguez2011}. As in the sphere-plane
geometry, the contributions from the different scattering channels are in
agreement with the results presented in Ref.~\cite{Milton2014} if the
identifications explained above are made.

In the high-temperature limit, $\nu\to\infty$, the polarization-conserving
channels yield a contribution to the Casimir free energy linear in temperature
while the contributions of the polarization-changing channels vanish. On the
other hand, all channels give rise to a negative Casimir free energy at
zero temperature. As a consequence, the contribution to the Casimir
entropy arising from the polarization-changing channels is negative for
all temperatures as already expected above on the basis of the matrix
elements of the translation operator.

From the expressions listed for the contributions to the Casimir free energy,
the corresponding contributions to the Casimir entropy can be obtained from its
definition (\ref{eq:defEntropy}). Introducing a rescaled Casimir entropy $s$ by
means of
\begin{equation}
\label{eq:rescaledEntropy_ss}
S = k_B\left(\frac{R_1R_2}{d^2}\right)^3s\,,
\end{equation}
one finds together with the abbreviation (\ref{eq:def_g})
\begin{equation}
\label{eq:ss_s0tmtm}
\begin{aligned}
s^{(0)}_{TM,TM} &= \frac{1}{\nu}\left[2g(\nu)\cosh(\nu)+2g(\nu)^2
                   -g(\nu)^3\cosh(\nu)\right.\\
&\qquad\quad\left.-g(\nu)^4\big(2\cosh^2(\nu)+1\big)\right]\\
&= 4s^{(0)}_{TE,TE}
\end{aligned}
\end{equation}
\begin{equation}
\label{eq:ss_s1tmtm}
\begin{aligned}
s^{(1)}_{TM,TM} &= \frac{1}{2\nu}\left[2g(\nu)\cosh(\nu)+2g(\nu)^2
                   +5g(\nu)^3\cosh(\nu)\right.\\
&\qquad\quad+g(\nu)^4\big(2\cosh^2(\nu)+1\big)\\
&\qquad\quad+g(\nu)^5\cosh(\nu)\big(\cosh^2(\nu)+2\big)\\
&\qquad\qquad\left.-g(\nu)^6\big(2\cosh^4(\nu)+11\cosh^2(\nu)+2\big)\right]\\
&= 4s^{(1)}_{TE,TE}
\end{aligned}
\end{equation}
\begin{equation}
\label{eq:ss_s1tmte}
\begin{aligned}
s^{(1)}_{TM,TE} &= \frac{1}{4\nu}\left[3g(\nu)^3\cosh(\nu)
+3g(\nu)^4\big(2\cosh^2(\nu)+1\big)\right.\\
&\qquad\quad+g(\nu)^5\cosh(\nu)\big(\cosh^2(\nu)+2\big)\\
&\qquad\quad\left.-g(\nu)^6\big(2\cosh^4(\nu)+11\cosh^2(\nu)+2\big)\right]\\
&= s^{(0)}_{TE,TM}\,.
\end{aligned}
\end{equation}

\begin{figure}
 \begin{center}
  \includegraphics[width=\columnwidth]{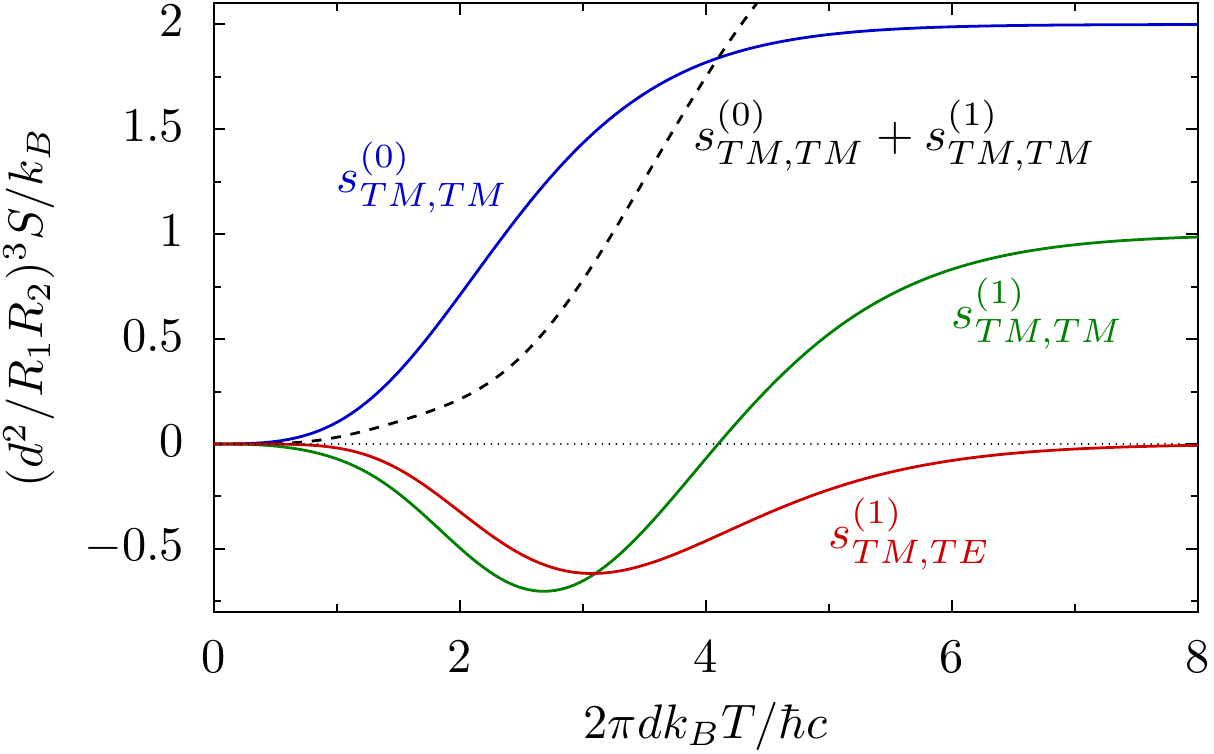}
 \end{center}
 \caption{The contributions (\ref{eq:ss_s0tmtm}), (\ref{eq:ss_s1tmtm}), and
 (\ref{eq:ss_s1tmte}) to the Casimir entropy of a sphere-sphere configuration
 are depicted as a function of the dimensionless temperature. While 
 $S^{(0)}_{TM,TM}$ is positive for all temperatures and $S^{(1)}_{TM,TM}$
 can be positive or negative depending on temperature, $S^{(1)}_{TM,TE}$ always
 yields a negative contribution to the Casimir entropy. The dashed line
 represents the sum of the contributions of the two polarization-conserving
 transverse magnetic channels.}
 \label{fig:ss_entropy_largedist}
\end{figure}

The temperature dependence of the contributions
(\ref{eq:ss_s0tmtm})--(\ref{eq:ss_s1tmte}) to the Casimir entropy is shown in
Fig.~\ref{fig:ss_entropy_largedist}. At first sight, the curves resemble those
presented in Fig.~\ref{fig:sp_entropy_largedist}. In both cases, the
contribution for $m=0$ is positive for all temperatures while the
polarization-conserving contribution for $m=1$ starts out negative and becomes
positive at higher temperatures. The polarization-changing channels always
yield a negative contribution to the Casimir entropy. A closer look reveals
that, in contrast to the plane-sphere configuration, the negative contribution
of the polarization-conserving channel with $m=1$ at low temperatures is much
bigger than the contribution of the polarization-changing channel. However, as
the dashed curve in Fig.~\ref{fig:ss_entropy_largedist} shows, the sum of the
Casimir entropies of the polarization-conserving channels remains positive at
all temperatures.

In order to analyze in more detail the negative contributions to the Casimir
entropy, we consider the low-temperature expansions of the expressions
(\ref{eq:ss_s0tmtm})--(\ref{eq:ss_s1tmte}) which are given by
\begin{align}
s^{(0)}_{TM,TM} &= \frac{4}{45}\nu^3+\frac{8}{315}\nu^5-\frac{8}{525}\nu^7
                   +O(\nu^9)\label{eq:ss_s0tmtm_lowt}\\
s^{(1)}_{TM,TM} &= -\frac{4}{45}\nu^3+\frac{2}{45}\nu^5-\frac{68}{1575}\nu^7
                   +O(\nu^9)\label{eq:ss_s1tmtm_lowt}\\
s^{(1)}_{TM,TE} &= -\frac{1}{63}\nu^5-\frac{2}{225}\nu^7+O(\nu^9)
\label{eq:ss_s1tmte_lowt}\,.
\end{align}
The dominant low-temperature contributions of order $T^3$ arise in the
polarization-conserving channels. However, they cancel each other. In view of
the fact that according to (\ref{eq:sp_s_lowtemp}), the Casimir entropy in the
plane-sphere configuration contains a leading term of order $T^3$, this may
come as a surprise. Indeed, a leading cubic term in the temperature can be
obtained for anisotropic objects \cite{Milton2014}. For the isotropic case
considered here, it can be shown that terms of order $\xi^5$ in the Mie
coefficients (\ref{eq:mie_a1}) and (\ref{eq:mie_b1}) lead to such a $T^3$ term
in the sphere-sphere configuration. However, this term is suppressed by a
factor of $(R/d)^3$ relative to the terms discussed here and therefore
negligible in the large-distance limit.

With the order $T^3$ not contributing to the entropy, the appearance of a
negative Casimir entropy is a nonperturbative effect \cite{Milton2014}. In the
next order, $T^5$, the negative contribution of the polarization-changing
channel is not sufficiently strong as compared to the positive contributions of
that order.  Therefore, at very low temperatures, the Casimir entropy of two
perfectly conducting spheres will be positive. On the other hand, it turns out
that the terms of order $T^7$, which in
(\ref{eq:ss_s0tmtm_lowt})--(\ref{eq:ss_s1tmte_lowt}) are all negative, lead
indeed to a negative Casimir entropy in an intermediate temperature range. This
was already shown in Ref.~\cite{Rodriguez2011} and is also visible in
Fig.~\ref{fig:ss_entropy_2d}.

It is now interesting to study how the different channels can contribute in
various ways to obtain either a positive Casimir entropy for all temperatures
or a negative Casimir entropy in a certain temperature regime. To this end, in
the next section we will also allow for spheres made of Drude-type metals.

\section{Perfectly conducting vs. Drude-type metal spheres}
\label{sec:drude}

So far, we have studied the behavior of the various scattering channels and
pointed out the relevance of the polarization-changing channels for the
appearance of a negative Casimir entropy. The weight with which the scattering
channels contribute can be modified by the physical properties of the objects
involved. In Ref.~\cite{Milton2014} this was done by choosing objects with
anisotropic polarizabilities and by varying their electric and magnetic
properties. Here, we will do the latter by allowing the objects to be either
made of perfectly conducting metals or Drude-type metals which have a finite
zero-frequency conductivity. The results will further underline the relevance
of the polarization-changing channels.

For simplicity, we will restrict ourselves in the following to setups
consisting of two spheres as depicted in Fig.~\ref{fig:geometry}b. Then, as
pointed out in Sect.~\ref{subsec:ss_large_distance}, the matrix elements of the
reflection operator are given by the Mie coefficients for $\ell=1$.  While for
perfectly conducting spheres, the Mie coefficients are the same up to a factor
-2, for Drude metal spheres with a dc conductivity $\sigma_0$, the reflection
of the TE mode can be neglected in the large-distance limit where in addition
to the conditions stated earlier, $d\gg\sigma_0 R^ 2/30c$ should hold.

These properties of the Mie coefficients allow us to construct three different
scenarios by choosing two spheres both made of perfect conductors, both made
of Drude metals, or one made of a perfect conductor and the other of a Drude
metal. In the first case, polarization-conserving as well as
polarization-changing channels contribute as discussed in
Sect.~\ref{subsec:ss_large_distance}. In contrast, in the second case, only modes
with TM polarization can complete roundtrips between the two spheres. Therefore,
in this case, the polarization-changing channel is completely suppressed.  In the
third case, only one of the two polarization-changing channels and its weight
relative to the polarization-conserving channels is modified with respect to two
perfectly conducting spheres.

Indicating in the superscript on the left-hand side the material of which the
two spheres are made, we obtain for the three situations just described the
rescaled entropy introduced in (\ref{eq:rescaledEntropy_ss}) with
\begin{align}
s^\text{PC/PC} &= \frac{5}{4}\left(s^{(0)}_{TM,TM}+s^{(1)}_{TM,TM}\right)
                  +2s^{(1)}_{TM,TE}\label{eq:s_pcpc}\\
s^\text{PC/D} &= s^{(0)}_{TM,TM}+s^{(1)}_{TM,TM}+s^{(1)}_{TM,TE}
                 \label{eq:s_pcd}\\
s^{D/D} &= s^{(0)}_{TM,TM}+s^{(1)}_{TM,TM}\label{eq:s_dd}\,,
\end{align}
where the components are given in (\ref{eq:ss_s0tmtm})--(\ref{eq:ss_s1tmte}).
These results are consistent with those obtained in Ref.~\cite{Milton2014}.

\begin{figure}
 \begin{center}
  \includegraphics[width=\columnwidth]{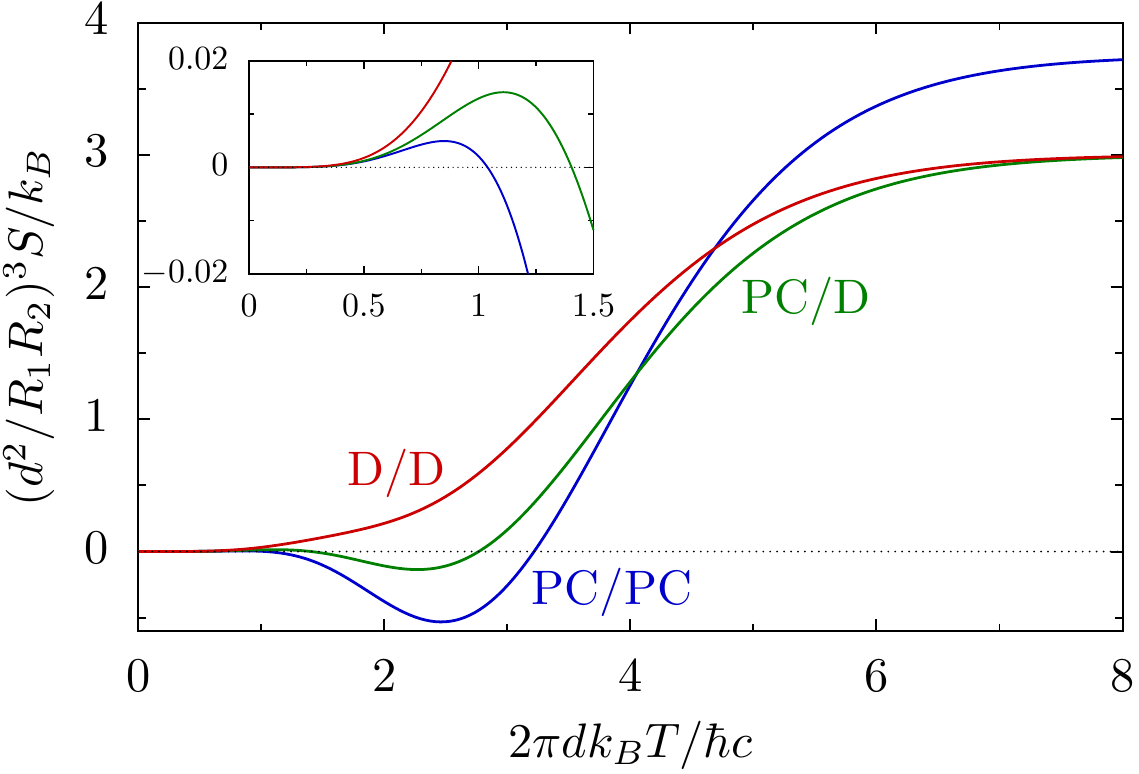}
 \end{center}
 \caption{Temperature dependence of the Casimir entropies
   (\ref{eq:s_pcpc})--(\ref{eq:s_dd}) for a sphere-sphere setup where the
   spheres are either made of perfect conductors (PC) or Drude metals (D).
   The inset shows the low-temperature behavior where a positive Casimir
   entropy is found for all three cases. The curve for D/D corresponds to
   the dashed curve in Fig.~\ref{fig:ss_entropy_largedist}.}
 \label{fig:ss_pc_d}
\end{figure}

The temperature dependence of the Casimir entropies
(\ref{eq:s_pcpc})--(\ref{eq:s_dd}) is displayed in Fig.~\ref{fig:ss_pc_d}.
At high temperatures, only the polarization-conserving channels contribute.
The difference between the case of perfectly conducting spheres and the other
two cases shows the suppression of the polarization-conserving TE channel due
to a Drude metal sphere.

The inset puts a special emphasis on the low-temperature behavior and shows
clearly that in all three cases, the Casimir entropy takes on positive values
for very low temperatures. However, only when all spheres are made of a Drude
metal does the Casimir entropy remain positive for all temperatures. This is the
case where no polarization-changing channels contribute.  

It is sufficient to allow for one polarization-changing channel by making
one of the spheres perfectly conducting in order to obtain a temperature
window in which the Casimir entropy becomes negative. The effect becomes
even more pronounced if both spheres are perfectly conducting as the weight
of the polarization-changing channels is doubled while the contribution
of the polarization-conserving channels is only increased by a quarter.

\section{Conclusions}

The origin of the negative Casimir entropy in the plane-sphere and sphere-sphere
configuration has been analyzed in the limit where the distance between the
objects is much larger than the radius of the sphere(s). In this limit, the
Casimir free energy and the Casimir entropy can easily be decomposed into the
contributions from various channels describing a roundtrip between the objects.
Three kinds of channels have been identified, differing significantly in the
temperature dependence of their contribution to the Casimir entropy.

The first kind of channels always makes a positive contribution to the Casimir
entropy. This was found to be the case for polarization-conserving channels
describing spherical waves with $m=0$.

The second kind of channels also conserves polarization but involves spherical
waves with $m=1$. While these channels yield a positive contribution to the
Casimir entropy at high temperatures, their contribution at sufficiently low
temperatures is negative. However, this negative part is compensated by the
polarization-conserving channels with $m=0$.

The third kind of channels is the most interesting one, because its contribution
to the Casimir entropy is negative for all temperatures. This behavior is 
associated with polarization-changing channels which exist for $m=1$ but not
for $m=0$. These channels are special, because their Casimir free energy
vanishes in the high-temperature limit. This fact and the ensuing negative
contribution to the Casimir entropy has been traced back to the 
polarization-changing nature of the channels. It can thus be concluded that 
polarization mixing in a scattering process is a crucial ingredient for the
appearance of a negative Casimir entropy, at least in the plane-sphere and
sphere-sphere configuration.

Which of the various channels contribute to the Casimir entropy can be influenced
by appropriately choosing the material out of which the scattering objects are
made. One can use the fact that in the long-distance limit the reflection of
transverse electric modes at spheres made of a Drude metal becomes negligible.
We have shown that for two Drude metal spheres, the Casimir entropy is positive
for all temperatures. In this situation, only roundtrip scattering processes
involving transverse magnetic modes are relevant. As soon as at least one of
the spheres is perfectly conducting, polarization-changing processes occur and
the Casimir entropy is found to become negative in a certain temperature
window.

\begin{acknowledgments}
KAM and GLI thank the Laboratoire Kastler Brossel for their hospitality during
the period of this work and CNRS and ENS for financial support. KAM's work was
further supported in part by grants from the Simons Foundation and the Julian
Schwinger Foundation.
\end{acknowledgments}

\end{document}